\documentclass[%
reprint,
superscriptaddress,
amsmath,amssymb,
aps,
prl,
longbibliography,
]{revtex4-1}

\usepackage{siunitx}

\DeclareSIUnit \mm {\milli\meter}
\DeclareSIUnit \cm {\centi\meter}
\DeclareSIUnit \us {\micro\second}
\DeclareSIUnit \ms {\milli\second}
\DeclareSIUnit \pA {\pico\ampere}
\DeclareSIUnit \pC {\pico\coulomb}
\DeclareSIUnit \fC {\femto\coulomb}
\DeclareSIUnit \fF {\femto\farrad}
\DeclareSIUnit \pF {\pico\farrad}
\DeclareSIUnit \mV {\milli\volt}
\DeclareSIUnit \kV {\kilo\volt}
\DeclareSIUnit \V {\volt}
\DeclareSIUnit \GOhm {\giga\ohm}
\DeclareSIUnit \MOhm {\mega\ohm}
\DeclareSIUnit \ton {\tonne}
\DeclareSIUnit \kton {\kilo\tonne}
\DeclareSIUnit \kt {\kilo\tonne}
\DeclareSIUnit \Mt {\mega\tonne}
\DeclareSIUnit \eV {\electronvolt}
\DeclareSIUnit \keV {\kilo\electronvolt}
\DeclareSIUnit \MeV {\mega\electronvolt}
\DeclareSIUnit \GeV {\giga\electronvolt}
\DeclareSIUnit \km {\kilo\meter}
\DeclareSIUnit \kW {\kilo\watt}
\DeclareSIUnit \MW {\mega\watt}
\DeclareSIUnit \MHz {\mega\hertz}
\DeclareSIUnit \kHz {\kilo\hertz}
\DeclareSIUnit \mrad {\milli\radian}
\DeclareSIUnit \year {year}
\DeclareSIUnit \POT {POT}
\DeclareSIUnit \sig {$\sigma$}
\DeclareSIUnit\parsec{pc}
\DeclareSIUnit\lightyear{ly}
\DeclareSIUnit\foot{ft}
\DeclareSIUnit\ft{ft}

\usepackage[printwatermark]{xwatermark}
\usepackage{graphicx}
\usepackage{dcolumn}
\usepackage{bm}
\usepackage{xcolor}
\usepackage{lineno}
\usepackage{mathtools}

\begin{document}
	
	\title{High-performance Generic Neutrino Detection in a LArTPC near the Earth's Surface with the MicroBooNE Detector}
	
%




\newcommand{\Bern}{Universit{\"a}t Bern, Bern CH-3012, Switzerland}
\newcommand{\BNL}{Brookhaven National Laboratory (BNL), Upton, NY, 11973, USA}
\newcommand{\UCSB}{University of California, Santa Barbara, CA, 93106, USA}
\newcommand{\Cambridge}{University of Cambridge, Cambridge CB3 0HE, United Kingdom}
\newcommand{\StKates}{St. Catherine University, Saint Paul, MN 55105, USA}
\newcommand{\CIEMAT}{Centro de Investigaciones Energ\'{e}ticas, Medioambientales y Tecnol\'{o}gicas (CIEMAT), Madrid E-28040, Spain}
\newcommand{\Chicago}{University of Chicago, Chicago, IL, 60637, USA}
\newcommand{\Cincinnati}{University of Cincinnati, Cincinnati, OH, 45221, USA}
\newcommand{\CSU}{Colorado State University, Fort Collins, CO, 80523, USA}
\newcommand{\Columbia}{Columbia University, New York, NY, 10027, USA}
\newcommand{\FNAL}{Fermi National Accelerator Laboratory (FNAL), Batavia, IL 60510, USA}
\newcommand{\Granada}{Universidad de Granada, Granada E-18071, Spain}
\newcommand{\Harvard}{Harvard University, Cambridge, MA 02138, USA}
\newcommand{\IIT}{Illinois Institute of Technology (IIT), Chicago, IL 60616, USA}
\newcommand{\KSU}{Kansas State University (KSU), Manhattan, KS, 66506, USA}
\newcommand{\Lancaster}{Lancaster University, Lancaster LA1 4YW, United Kingdom}
\newcommand{\LANL}{Los Alamos National Laboratory (LANL), Los Alamos, NM, 87545, USA}
\newcommand{\Manchester}{The University of Manchester, Manchester M13 9PL, United Kingdom}
\newcommand{\MIT}{Massachusetts Institute of Technology (MIT), Cambridge, MA, 02139, USA}
\newcommand{\Michigan}{University of Michigan, Ann Arbor, MI, 48109, USA}
\newcommand{\Minnesota}{University of Minnesota, Minneapolis, MN, 55455, USA}
\newcommand{\NMSU}{New Mexico State University (NMSU), Las Cruces, NM, 88003, USA}
\newcommand{\Otterbein}{Otterbein University, Westerville, OH, 43081, USA}
\newcommand{\Oxford}{University of Oxford, Oxford OX1 3RH, United Kingdom}
\newcommand{\PNNL}{Pacific Northwest National Laboratory (PNNL), Richland, WA, 99352, USA}
\newcommand{\Pitt}{University of Pittsburgh, Pittsburgh, PA, 15260, USA}
\newcommand{\Rutgers}{Rutgers University, Piscataway, NJ, 08854, USA}
\newcommand{\StMarys}{Saint Mary's University of Minnesota, Winona, MN, 55987, USA}
\newcommand{\SLAC}{SLAC National Accelerator Laboratory, Menlo Park, CA, 94025, USA}
\newcommand{\SDSMT}{South Dakota School of Mines and Technology (SDSMT), Rapid City, SD, 57701, USA}
\newcommand{\Maine}{University of Southern Maine, Portland, ME, 04104, USA}
\newcommand{\Syracuse}{Syracuse University, Syracuse, NY, 13244, USA}
\newcommand{\TelAviv}{Tel Aviv University, Tel Aviv, Israel, 69978}
\newcommand{\Tennessee}{University of Tennessee, Knoxville, TN, 37996, USA}
\newcommand{\UTA}{University of Texas, Arlington, TX, 76019, USA}
\newcommand{\Tufts}{Tufts University, Medford, MA, 02155, USA}
\newcommand{\VTech}{Center for Neutrino Physics, Virginia Tech, Blacksburg, VA, 24061, USA}
\newcommand{\Warwick}{University of Warwick, Coventry CV4 7AL, United Kingdom}
\newcommand{\Yale}{Wright Laboratory, Department of Physics, Yale University, New Haven, CT, 06520, USA}

\affiliation{\Bern}
\affiliation{\BNL}
\affiliation{\UCSB}
\affiliation{\Cambridge}
\affiliation{\StKates}
\affiliation{\CIEMAT}
\affiliation{\Chicago}
\affiliation{\Cincinnati}
\affiliation{\CSU}
\affiliation{\Columbia}
\affiliation{\FNAL}
\affiliation{\Granada}
\affiliation{\Harvard}
\affiliation{\IIT}
\affiliation{\KSU}
\affiliation{\Lancaster}
\affiliation{\LANL}
\affiliation{\Manchester}
\affiliation{\MIT}
\affiliation{\Michigan}
\affiliation{\Minnesota}
\affiliation{\NMSU}
\affiliation{\Otterbein}
\affiliation{\Oxford}
\affiliation{\PNNL}
\affiliation{\Pitt}
\affiliation{\Rutgers}
\affiliation{\StMarys}
\affiliation{\SLAC}
\affiliation{\SDSMT}
\affiliation{\Maine}
\affiliation{\Syracuse}
\affiliation{\TelAviv}
\affiliation{\Tennessee}
\affiliation{\UTA}
\affiliation{\Tufts}
\affiliation{\VTech}
\affiliation{\Warwick}
\affiliation{\Yale}

\author{P.~Abratenko} \affiliation{\Tufts} 
\author{M.~Alrashed} \affiliation{\KSU}
\author{R.~An} \affiliation{\IIT}
\author{J.~Anthony} \affiliation{\Cambridge}
\author{J.~Asaadi} \affiliation{\UTA}
\author{A.~Ashkenazi} \affiliation{\MIT}
\author{S.~Balasubramanian} \affiliation{\Yale}
\author{B.~Baller} \affiliation{\FNAL}
\author{C.~Barnes} \affiliation{\Michigan}
\author{G.~Barr} \affiliation{\Oxford}
% ...

\collaboration{The MicroBooNE Collaboration}
\thanks{microboone\_info@fnal.gov}\noaffiliation



	\date{\today}
	
	\begin{abstract}
		Large Liquid Argon Time Projection Chambers (LArTPCs) are being increasingly adopted in neutrino oscillation experiments because of their superb imaging capabilities through the combination of both tracking and calorimetry in a fully active volume. Active LArTPC neutrino detectors at or near the Earth's surface, such as the MicroBooNE experiment, present a unique analysis challenge because of the large flux of cosmic-ray muons and the slow drift of ionization electrons. We present a novel Wire-Cell-based high-performance generic neutrino-detection technique implemented in MicroBooNE. The cosmic-ray background is reduced by a factor of 1.4$\times10^{5}$ resulting in a 9.7\% cosmic contamination in the selected neutrino candidate events, for visible energies greater than 200~MeV, while the neutrino signal efficiency is retained at 88.4\% for $\nu_{\mu}$ charged-current interactions in the fiducial volume in the same energy region. This significantly improved performance compared to existing reconstruction algorithms, marks a major milestone toward reaching the scientific goals of LArTPC neutrino oscillation experiments operating near the Earth's surface.
	\end{abstract}
	
	\maketitle
	
	
    The Liquid Argon Time Projection Chamber (LArTPC)~\cite{rubbia77,Chen:1976pp,willis74,Nygren:1976fe} is an advanced technology to detect neutrinos with its superb imaging capabilities through the combination of both tracking and calorimetry in a fully active volume. Such capabilities make LArTPC detectors attractive to address important unresolved questions in neutrino physics, such as the presence of CP violation in the lepton sector~\cite{nova,t2k}, the order of neutrino
    masses~\cite{Qian:2015waa}, the existence of sterile neutrinos~\cite{Machado:2019oxb} as well as the precise measurements of neutrino-nucleus interactions~\cite{pdg-nn}.
	In the past two decades, the LArTPC technology has gone through rapid development~\cite{Cavanna:2018yfk}, and detectors with active mass ranging up to 500~tons have been constructed and operated~\cite{Amerio:2004ze,ArgoNeuT2012,Acciarri:2016smi,Badhrees:2012zz,Bhandari:2019rat,Hahn:2016tia,Cavanna:2014iqa,Abi:2020mwi}.
	Looking forward, the Short-Baseline Neutrino program (SBN)~\cite{Antonello:2015lea}
	is under construction and partially in operation with the main goal of resolving a class of experimental anomalies in neutrino physics to which the existence of light sterile neutrinos is a possible explanation. Going further, the Deep Underground Neutrino Experiment (DUNE), with multiple 10~kton active mass LArTPC modules as the far detector, is a next generation long-baseline neutrino oscillation experiment aiming to reveal new symmetries of nature~\cite{dune-tdr-1}. To ensure the success of these future physics programs, the current-generation large LArTPCs, including MicroBooNE~\cite{Acciarri:2016smi} and the ProtoDUNEs~\cite{Abi:2020mwi}, are critical to develop and demonstrate the full capability of the LArTPC technology.

	MicroBooNE experiment locates at the Booster Neutrino Beam (BNB)~\cite{AguilarArevalo:2008yp} at the Fermi National Accelerator Laboratory in Batavia, IL, USA.
Inside a single-walled cryostat with a 170 ton capacity, the detector~\cite{Acciarri:2016smi} consists of a $\SI{2.56}{\meter}$  $\times$ $\SI{2.32}{\meter}$ $\times$ $\SI{10.36}{\meter}$ active TPC (85 metric tons)
for ionization charge detection, and an array of 32 photomultiplier tubes (PMTs)~\cite{Briese:2013wua} for scintillation light detection. The anode consists of three parallel wire-readout planes, spaced 3 mm apart, with each plane's wires rotated by 60 degrees relative to the other planes and with a 3~mm wire pitch, for a total number of 8256 wires. Ionization electrons from charged particles drift through the LAr drift toward the wire planes at the anode. 
The drift speed at the operating electric field of \SI{273}{\V/\cm} is \SI{1.1}{\mm/\us}, leading to a 2.3~ms drift time for the maximum 2.56~m drift distance. The induced current on the wires is amplified and shaped through the custom-designed analog front-end electronics readout~\cite{Radeka:2011zz} operating at 89~K in the LAr.
	Data recording of candidate neutrino interactions is triggered on a hardware level by each BNB beam spill (within a 1.6~$\mu$s time window). Because of the slow drift of ionization electrons, both TPC and PMT readout windows are extended relative to the beam spill to include both neutrino interactions and spill-in cosmic-ray muon background. For the TPC readout, a digitized waveform with 9600 samples at a 2~MHz sampling rate, spanning from -1.6~ms to +3.2~ms relative to the trigger, is recorded for each wire. For the PMT readout, a digitized waveform with 1500 samples at a 64~MHz sampling rate covering the beam spill is recorded for each PMT. In addition, self-discriminated readout waveforms, each with 40 samples, are taken during a period of 6.4~ms around the BNB trigger to record cosmic-rays nearby in time. At the nominal BNB intensity of approximately 4$\times$10$^{12}$ protons on target (POT) per spill, one neutrino interaction is expected inside the TPC 
	active volume every 600 spills.
	
	The main challenge to detect neutrinos comes from the large cosmic-ray background because of the near-surface location of the detector. During data acquisition, a software selection requiring PMT signals in coincidence with a beam spill to exceed a certain threshold of photoelectrons (PEs) is applied, and this reduces the number of recorded events by a factor of 22. Nonetheless, over 95\% of the remaining events are still triggered by cosmic rays in coincidence with or arriving just before the beam spill. In addition, at the 5.5~kHz cosmic-ray rate~\cite{Acciarri:2017rnj}, there are on average 26 cosmic-ray muons in the 4.8~ms TPC readout window. 
	This large cosmic-ray background compromises both the purity and efficiency performance in selecting neutrino interactions, as noted in prior work~\cite{Adams:2018fud,Adams:2018sgn,Adams:2018lzd,Adams:2019iqc}.
	In this letter, a new analysis procedure with a significantly improved performance of cosmic-ray background rejection while maintaining a high efficiency in detecting neutrinos is described. The procedure consists of a series of newly developed event processing and reconstruction techniques, including offline light reconstruction to reject events triggered by cosmic rays that arrive just before the beam spill, charge-light matching to remove cosmic-rays outside the beam spill, rejection of through-going cosmic-ray muons based on geometry information, rejection of stopped cosmic-ray muons based on calorimetry information, and rejection of events with incorrect charge-light matching. 
	
	PMT waveforms are processed offline to reconstruct flashes,  clusters of PMT signals occurring close in time. 
    For the recorded waveforms triggered by the beam spill, a deconvolution procedure based on the Fast Fourier Transform is performed to remove the PMT readout response. 
	A flash is then formed by requiring that more than two PMTs record activity of greater than 1.5 PEs, and the total number of PEs recorded by sum of all PMTs is greater than 6 within 100~ns. 
    Unless another flash is found, the time window for a flash lasts 7.2~$\mu$s to include the contribution from the slow component of the scintillation light. 
    Similarly, flashes are formed for the self-discriminated waveforms outside the beam spill, where the number of PEs is directly derived from the integral of the waveform after taking into account the single PE response and an extrapolation of the amount of late scintillation light outside the readout window.
    About 70\% of the cosmic rays that pass the software trigger are rejected by requiring at least one reconstructed flash to coincide in time with the beam spill. The rejected events are mainly triggered by the late scintillation light from cosmic-ray muons arriving just before the beam spill.
	
	The TPC data processing procedure includes excess noise removal~\cite{Acciarri:2017sde} and signal processing~\cite{Adams:2018dra}. 
	The signal processing procedure adopts a novel 2D deconvolution technique~\cite{Adams:2018dra} to extract the ionization charge distribution, which significantly improves the performance for induction wire planes in comparison with a one-dimensional (1D) deconvolution technique~\cite{Baller:2017ugz} used in previous work. 
	Using this approach, a good agreement between the observed~\cite{Adams:2018gbi} and simulated~\cite{Adams:2018dra} reconstructed charge in all three wire planes has been achieved.
	
	The 2D charge measurements from wire planes are fed into a tomographic three-dimensional (3D) image reconstruction algorithm, Wire-Cell~\cite{Qian:2018qbv}, in which a 
	cross-sectional image in each 2~$\mu$s drift-time slice is reconstructed. First, overlapping areas of wire activity, blobs, are created using the wire geometry information. Then, spurious blobs are removed by utilizing the charge information. We use the generic constraints in Wire-Cell to reconstruct a 3D event image independent of its topology (e.g.~track or 
	electromagnetic shower). 
	In regions of the detector with non-functional wires~\cite{Acciarri:2017sde}, a special algorithm reconstructs blobs from activity in just two wire planes. 
	This reduces the unusable detector volume by a factor of 10 from 30\% to 3\%~\cite{wire-cell-uboone}.
	Additional algorithms such as iterative image reconstruction and clustering are implemented to further remove spurious blobs and to improve the quality of the 3D images~\cite{wire-cell-uboone}.
	
\begin{figure*}[htb]
		\includegraphics[width=0.55\textwidth]{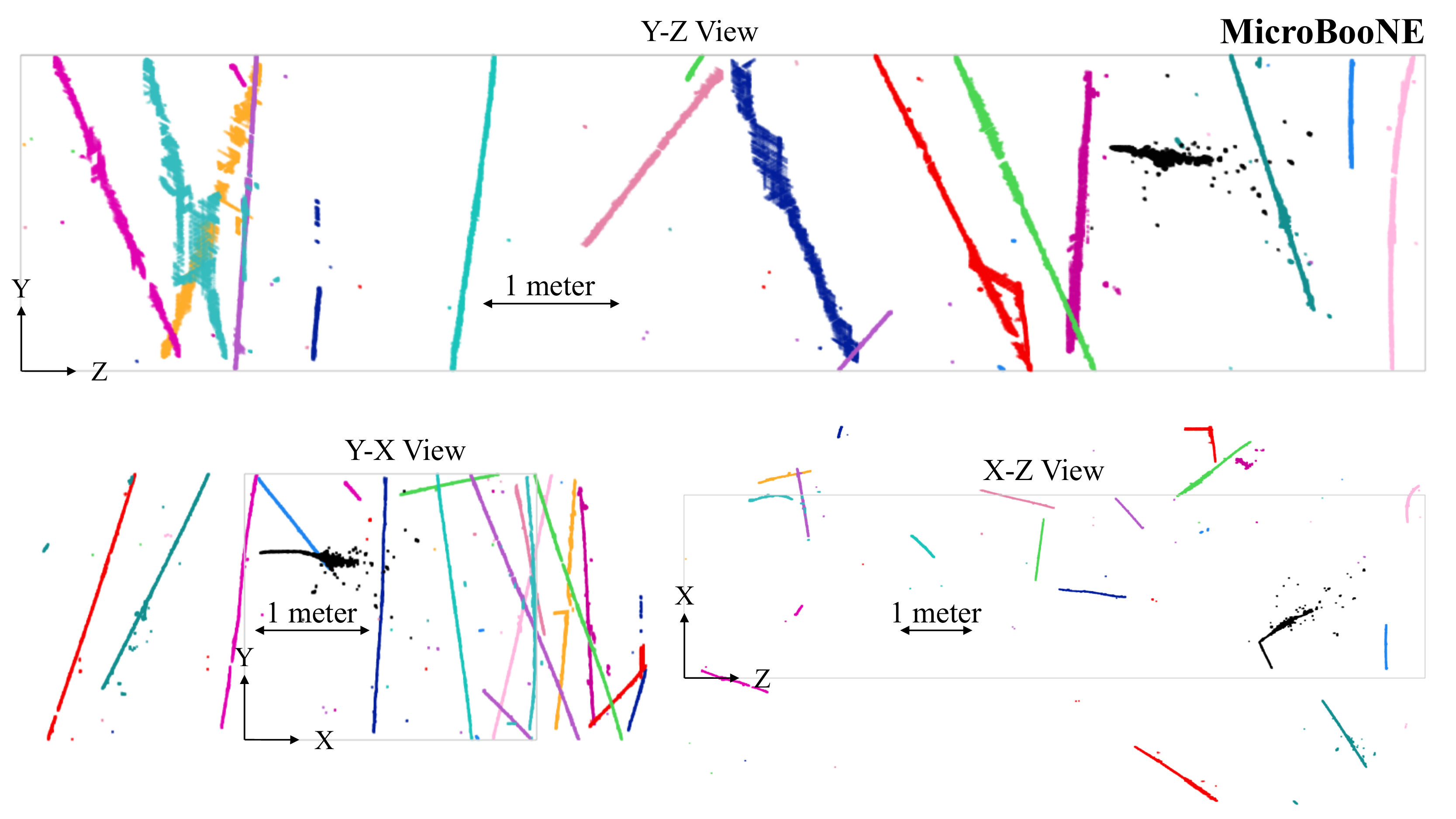}
		\includegraphics[width=0.44\textwidth]{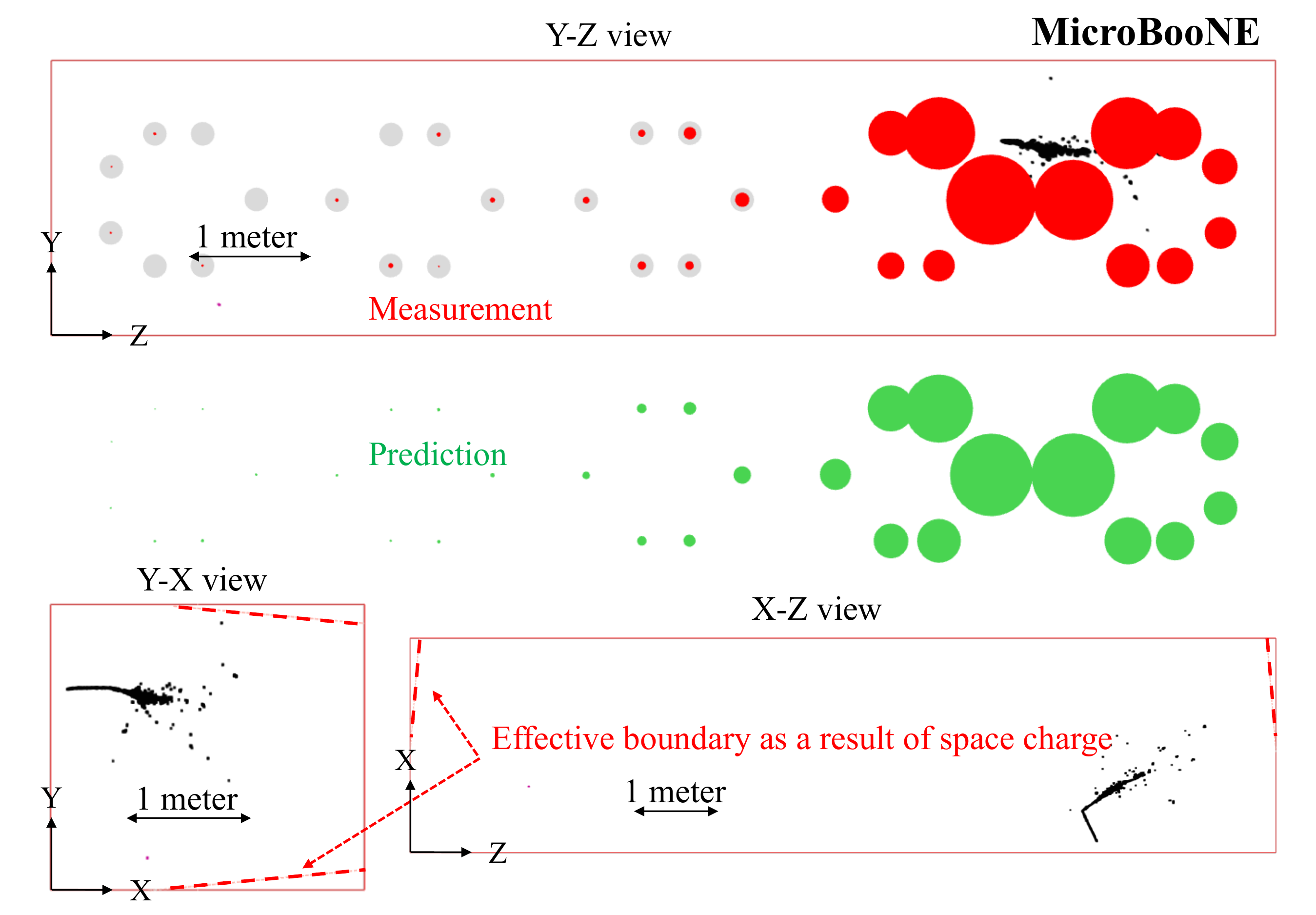}
		\put(-360,-10){(a)}
		\put(-140,-10){(b)}
		\caption{\label{fig:matching} A $\nu_e$ charged current (CC) interaction candidate from MicroBooNE data. The X axis is drift electric field direction from the TPC anode to the cathode. The Y axis is vertical up, and the Z axis is along the neutrino beam direction.
			Panel (a) shows three projections of reconstructed 3D clusters in the full TPC readout window before TPC-charge/PMT-light matching. Each cluster is shown in a different color. The gray box represents the TPC active volume while the two ends along the X axis represent the trigger time and the maximum drift time relative to the trigger.
			Panel (b) shows the projections of the $\nu_e$CC candidate cluster after applying the TPC-charge/PMT-light matching. The red (green) circles represent the observed (predicted) PEs at each PMT with their area proportional to the PE.
			The consistency between the measured and predicted light pattern indicates a good match. 
			The effective detector boundary as a result of space charge effects is indicated by the red dashed lines in the corner of the TPC active volume as shown in the ``Y-X view'' and ``X-Z view''.}
	\end{figure*}

	Typically, there are thousands of blobs in a reconstructed 3D event image. They are further grouped into clusters that represent individual physical signals from cosmic-ray muons or a neutrino interaction. 
	The 3D charge cluster from a single physical signal is identified using a set of algorithms based on connectivity and proximity~\cite{wire-cell-uboone}. 
	Special algorithms are implemented to mitigate gaps in the 3D image caused by the 3\% unusable volume, the  
	imperfect coherent excess noise removal~\cite{Acciarri:2017sde}, 
	and the inefficiency of signal processing for the prolonged track topology, i.e. tracks parallel to the drift direction~\cite{Adams:2018dra}. 
    On the other hand, over-clustering may occur when ionization charges produced at different time and drift distance but at the same projected position on the anode plane arrive at the anode plane at the same time, leading to two separated tracks identified as one cluster. An algorithm is created to separate tracks in this case. 
	Figure~\ref{fig:matching}a shows an event image after the 3D clustering. 
	
	One particular challenge in the event reconstruction for a LArTPC, compared to other types of tracking calorimeters such as those deployed in NO$\nu$A~\cite{Adamson:2016xxw}, 
	MINER$\nu$A~\cite{Aliaga:2013uqz}, and MINOS~\cite{Michael:2008bc}, is that the event topology 
	information from ionization charge and the timing information from scintillation light are decoupled because of the slow drift of ionization electrons. 
	In a typical BNB event, the number of TPC clusters in the TPC readout window is 20--30, and the number of PMT flashes is 40--50. The latter is larger because of the contribution from the LAr volume outside the active TPC.
	Since the scintillation light is detected by PMTs on a much shorter time scale, the PMT flashes can be used to distinguish each individual TPC signal and to provide its event time. This is especially useful to select in-beam neutrino activity.
	A novel many-to-many charge-light matching algorithm is used to find the corresponding PMT flash for each TPC charge cluster~\cite{wire-cell-uboone}. 
	All possible pairs of TPC clusters and PMT flashes are constructed after considering the allowed drift time window for each PMT flash. 
	With each hypothetical cluster-flash pair, the observed PMT light pattern can be compared to 
	the predicted pattern, assuming that the produced light is proportional to the reconstructed 3D charge. The prediction of light at each PMT also considers the 
	acceptance and propagation of the light as a function of the 3D position, which is parameterized by a photon library generated by Geant4~\cite{geant,geant_2}. 
	The compressed sensing technique~\cite{cs} used in the 3D image reconstruction is again implemented here to
	select the best hypotheses considering that one cluster can match to zero or one flash and one flash can match to zero, one, or multiple clusters. The average accuracy of the 
	charge-light matching is roughly 95\%, evaluated by both Monte Carlo (MC) simulation and data, the latter through a hand-scanning study. 
	Figure~\ref{fig:matching}b shows a data example where the single TPC cluster from a beam $\nu_e$ CC interaction candidate is selected after matching with an in-beam PMT flash. 
	After charge-light matching, the cosmic-ray muon background, which is mostly outside the beam spill window, is significantly reduced by a factor of 30-40. However, the majority of the remaining in-beam candidates still originate from cosmic-ray muons. Various techniques were developed to reject these cosmic backgrounds with limited impact on the efficiency for identifying neutrino interactions. A brief introduction on these techniques is as follows and detailed description can be found in Ref.~\cite{wire-cell-generic-neutrino}.
    
    The largest remaining background is through-going muons (TGMs) that coincides in time with the beam spill, and the number of TGMs is about 5 times larger than that of neutrino interactions in the active volume after charge-light matching.
    The identification of TGMs relies on a precise knowledge of the effective boundary of the TPC active volume. Because of the high rate of cosmic-ray muons traversing the detector volume, the accumulation of positively charged argon ions (i.e.~space charge) results in a distorted electric field, thus modifying the reconstructed position from the true position of ionization electrons~\cite{Adams:2019qrr, Abratenko:2020bbx}.  
	The effective boundary shown in Fig.~\ref{fig:matching}b is calibrated 
	with start-time-corrected positions of cosmic-ray muon clusters after charge-light matching. The fiducial boundary 
	used to identify TGMs is defined as 3~cm inside the effective boundary, which leads to a fiducial volume of 94.2\% of the active TPC~\cite{wire-cell-generic-neutrino}.
	
	The second largest background results from stopped muons (STMs), 
	which enter the active volume and stop inside. 
	STMs are identified based on their direction, which is determined by identifying an increase of ionization charge per unit length ($dQ/dx$) near the end of the track (i.e.~Bragg peak).
	Inspired by the projection matching algorithm~\cite{antonello:2012hu}, 
	a new 3D track trajectory and $dQ/dx$ fitting procedure was developed~\cite{wire-cell-generic-neutrino}. 
	First, an initial seed of the track trajectory is obtained by constructing a Steiner-tree graph~\cite{Steiner} from the 3D points in the cluster and finding the shortest paths between extreme points. 
	The Steiner-tree ensures that points associated with the largest charges are included in the initial seed.
	Then, the best-fit 3D trajectory points with about 6~mm spacing are obtained by minimizing a charge-weighted distance, which is constructed to compare the 2D measurements in time-versus-wire views from the three wire planes to the predictions given a 3D trajectory point.
	A numerical solver for large linear systems (BiCGSTAB~\cite{BiCGSTAB}) is utilized to perform a minimization. 
	With the trajectory determined, the $dQ/dx$ associated with each trajectory point can be obtained by minimizing the squared difference 
	between the reconstructed and predicted ionization charge. 
	A parameterized model is used to predict the measured charge taking into account the diffusion of ionization electrons during transportation and the smearing of the charge distribution in the signal processing.
	This two-step procedure is adopted to avoid a nonlinear fitting process, ensuring the stability of the fit. 
	Regularization on smoothness is included to further improve the $dQ/dx$ fitting performance. 
	

	The third largest background is categorized as light-mismatched (LM) events in which the measured light pattern is not consistent with the predicted pattern from the matched TPC clusters. The consistency is examined by a Kolmogorov-Smirnov test~\cite{ks_test1}.
	A dedicated LM tagger~\cite{wire-cell-generic-neutrino} is used to remove various LM events: low visible energy events with short track lengths or small predicted PE signal, mismatched events caused by the inefficiency of the light detection system for cathode-side events which have low light acceptance for PMTs, mismatched events caused by the incomplete prediction of light production originating from activities outside the TPC active volume,
	and incorrectly matched events that coincidentally have a reasonably good light pattern match.
	For LM events, both the light and topology information is used to seek a different TPC cluster that agrees with the measured light pattern and is consistent with a boundary-crossing muon. 
	The external cosmic-ray tagger~\cite{Adams:2019bzt} system may provide additional rejection of the LM events but is not included in this work. 
	
	\begin{table*}[t]
		\begin{center}
			\caption{\label{tab:table}
				Summary of the cosmic-ray background rejection power, the cumulative selection efficiency for neutrino interactions in the fiducial volume (94.2\% of the active volume), and the neutrino signal to the cosmic-ray background ratio for each selection criterion.
				The relative cosmic-ray rejection power to the previous criterion is shown in parentheses. 
                The numbers come from MC simulation of BNB neutrino interactions or beam-off data. 
                The errors are statistical only.  
				Neutrinos originating outside the fiducial volume are not counted in this table.
				See Fig.~\ref{fig:results} for more details of the selected neutrino candidates.}
			\begin{tabular}{ccccc}
				\hline
				\hline
                Selection & $\nu_{\mu}$ CC efficiency & $\nu_{\mu}$ NC efficiency & cosmic-ray reduction & $\nu_{\mu}$ : cosmic-ray \\\hline
				Hardware trigger     & 100\% & 100\% & 1 (1) & 1 : 20000\\
            Light filter & (98.31$\pm$0.03)\%  & (85.4$\pm$0.1)\%  & (0.998$\pm$0.002)$\times$10$^{-2}$ (0.01) & 1 : 210\\
            Charge-light matching & (92.1$\pm$0.1)\%  & (53.6$\pm$0.2)\%  & (2.62$\pm$0.04)$\times$10$^{-4}$ (0.026) & 1 : 6.4 \\
            Through-going muon rejection & (88.9$\pm$0.1)\%  & (52.1$\pm$0.2)\%  & (4.4$\pm$0.2)$\times$10$^{-5}$ (0.17) & 1.1 : 1\\
            Stopped muon rejection & (82.9$\pm$0.1)\%  & (50.3$\pm$0.2)\%  & (1.4$\pm$0.1)$\times$10$^{-5}$ (0.32) & 2.8 : 1\\
            Light-mismatch rejection  & (80.4$\pm$0.1)\%  & (35.9$\pm$0.2)\%  & (6.9$\pm$0.6)$\times$10$^{-6}$ (0.50) & 5.2 : 1\\
				\hline
				\hline
			\end{tabular}
		\end{center}
	\end{table*}
	
	\begin{figure}[htpb!]
		\includegraphics[width=0.48\textwidth]{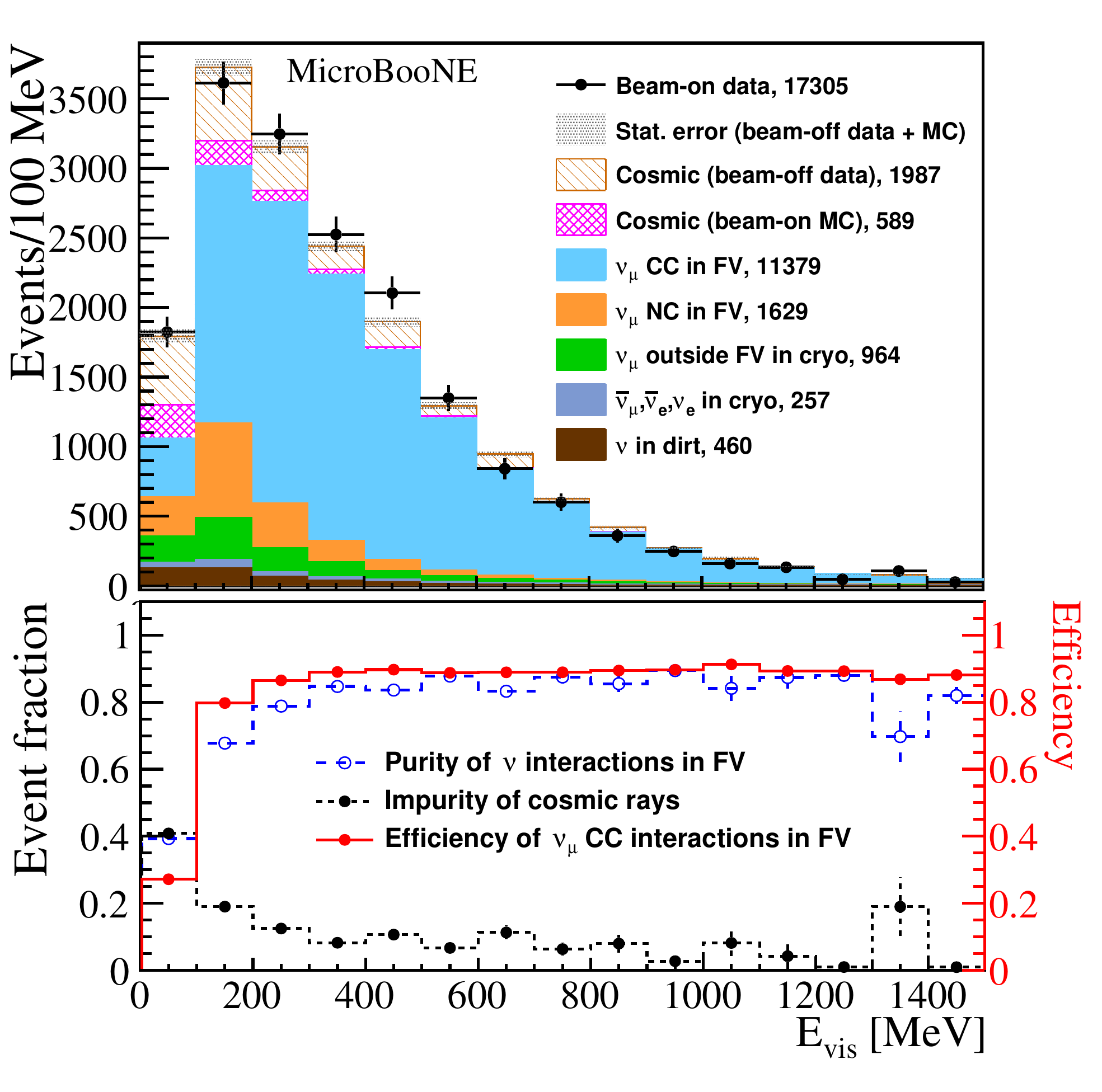}
		\caption{\label{fig:results} 
			Top: The selected events from the BNB beam-on data sample as a function of visible energy reconstructed in the detector are compared to the stacked selected events from beam-on MC simulation and beam-off data samples. All scaled to 5$\times10^{19}$ POT.
			Each event in the MC sample is guaranteed to have a neutrino interaction in the active volume or in the cryo.
			The selected neutrino events are categorized using the MC truth information.
			Bottom: The selection efficiency of $\nu_\mu$ CC events originating in the fiducial volume and the event fraction for neutrino signal or cosmic-ray background are shown. The dip/jump in ``purity'' and ``impurity'' around 1400 MeV is believed to be caused by the statistical fluctuation of cosmic-ray background events in that bin.}
	\end{figure}
	
	Table~\ref{tab:table} summarizes the performance of the cosmic-ray background rejection and the corresponding neutrino selection efficiency, where the ``light filter'' step combines both the software trigger and the offline flash reconstruction.
    An overall 1.4$\times$10$^{5}$ rejection factor is achieved, leading to a cosmic-ray impurity of 9.7\% (14.9\%) for reconstructed visible energy, E$_{\rm vis}$, greater than 200 (0) MeV as shown in Fig.~\ref{fig:results}. E$_{\rm vis}$ is converted from the total measured charge taking into account the recombination and attenuation of the ionization electrons~\cite{wire-cell-generic-neutrino}.  
	In addition, these algorithms retain a high fraction of the neutrino interactions originating in the fiducial volume with 88.4\% (80.4\%) of $\nu_\mu$ charged current (CC) neutrino interactions and 80.8\% (35.9\%) $\nu_\mu$ neutral current (NC) neutrino interactions remaining for E$_{\rm vis}$ greater than 200 (0) MeV. 
    The loss of NC interactions below 200 MeV is due to the large fraction of NC interactions without easily reconstructable low energy hadronic final state.
	
	The selected events from beam-on MC simulation and beam-off data samples are compared to the selected events from a beam-on data sample, as shown in Fig.~\ref{fig:results}. 
	All reported numbers are scaled to a BNB exposure of 5$\times$10$^{19}$ POT. The error bars are statistical only.
	The selected simulated neutrino events are categorized based on their interaction types and locations, including inside the fiducial volume (FV), inside the nonfunctional LAr volume (cryo), and outside the LAr volume (dirt).
	The main cosmic-ray background that coincides in time with the beam spill is estimated from the beam-off data sample.
	An additional cosmic-ray background corresponds to a cosmic-ray cluster incorrectly matched to the neutrino-induced PMT flash. 
	
	In summary, the work presented in this letter using strictly TPC and PMT information marks a major milestone toward fully achieving the scientific goals of LArTPC neutrino oscillation experiments operating near the surface. The performance of the cosmic-ray background rejection and the generic neutrino selection efficiency is significantly improved compared to previous 
	results~\cite{Adams:2018fud,Adams:2018sgn,Adams:2018lzd,Adams:2019iqc}.
   In this work, the overall selection efficiency of inclusive $\nu_\mu$ CC events in the fiducial volume is 80.4\% (88.4\%), with an overall cosmic contamination of 14.9\% (9.7\%), for visible energies greater than 0 (200) MeV. About 99.98\% of cosmic-ray backgrounds are rejected after software triggering.
    Compared to the result in Ref.~\cite{Adams:2019iqc}, the overall selection efficiency of inclusive $\nu_\mu$ CC events in the TPC active volume is increased by a factor of 2.7, with an enlargement of the fiducial volume by a factor of 1.9 and a reduction of the cosmic contamination by a factor of 2.4. Meanwhile, only about 10\% $\nu_e$ CC events are rejected~\cite{wire-cell-generic-neutrino}.
    Since many LArTPC based neutrino oscillation experiments will be statistics limited the work presented here describes a technique which significantly increases sample sizes therefore improving the sensitivity, precision and effectiveness of these detectors. 
    The generic neutrino detection forms a solid
    foundation for a high-performance $\nu_e$CC and $\nu_\mu$CC event 
    selection allowing for a compelling test of 
    $\nu_e$ explanation of MiniBooNE low-energy excess~\cite{Aguilar-Arevalo:2020nvw} and high-precision multi-fold 
    differential charged-current $\nu_\mu$-Ar cross section measurements.
    Looking forward, many of the novel techniques summarized in this work 
    can be naturally adopted into and expected to have a significant performance impact on the SBN~\cite{Antonello:2015lea} and DUNE~\cite{dune-tdr-1} experiments.
%
	
	\acknowledgments
	This document was prepared by the MicroBooNE collaboration using the
	resources of Fermi National Accelerator Laboratory (Fermilab), a
	U.S. Department of Energy, Office of Science, HEP User Facility.
	Fermilab is managed by Fermi Research Alliance, LLC (FRA), acting
	under Contract No. DE-AC02-07CH11359.  MicroBooNE is supported by the
	following: the U.S. Department of Energy, Office of Science, Offices
	of High Energy Physics and Nuclear Physics; the U.S. National Science
	Foundation; the Swiss National Science Foundation; the Science and
	Technology Facilities Council (STFC), part of the United Kingdom 
	Research and Innovation; and The Royal Society (United Kingdom).  
	Additional support for the laser calibration system and cosmic-ray tagger was provided by the Albert Einstein Center for Fundamental Physics, Bern, Switzerland.
	
	\bibliography{main}

\begin{thebibliography}{49}%
\makeatletter
\providecommand \@ifxundefined [1]{%
 \@ifx{#1\undefined}
}%
\providecommand \@ifnum [1]{%
 \ifnum #1\expandafter \@firstoftwo
 \else \expandafter \@secondoftwo
 \fi
}%
\providecommand \@ifx [1]{%
 \ifx #1\expandafter \@firstoftwo
 \else \expandafter \@secondoftwo
 \fi
}%
\providecommand \natexlab [1]{#1}%
\providecommand \enquote  [1]{``#1''}%
\providecommand \bibnamefont  [1]{#1}%
\providecommand \bibfnamefont [1]{#1}%
\providecommand \citenamefont [1]{#1}%
\providecommand \href@noop [0]{\@secondoftwo}%
\providecommand \href [0]{\begingroup \@sanitize@url \@href}%
\providecommand \@href[1]{\@@startlink{#1}\@@href}%
\providecommand \@@href[1]{\endgroup#1\@@endlink}%
\providecommand \@sanitize@url [0]{\catcode `\\12\catcode `\$12\catcode
  `\&12\catcode `\#12\catcode `\^12\catcode `\_12\catcode `\%12\relax}%
\providecommand \@@startlink[1]{}%
\providecommand \@@endlink[0]{}%
\providecommand \url  [0]{\begingroup\@sanitize@url \@url }%
\providecommand \@url [1]{\endgroup\@href {#1}{\urlprefix }}%
\providecommand \urlprefix  [0]{URL }%
\providecommand \Eprint [0]{\href }%
\providecommand \doibase [0]{http://dx.doi.org/}%
\providecommand \selectlanguage [0]{\@gobble}%
\providecommand \bibinfo  [0]{\@secondoftwo}%
\providecommand \bibfield  [0]{\@secondoftwo}%
\providecommand \translation [1]{[#1]}%
\providecommand \BibitemOpen [0]{}%
\providecommand \bibitemStop [0]{}%
\providecommand \bibitemNoStop [0]{.\EOS\space}%
\providecommand \EOS [0]{\spacefactor3000\relax}%
\providecommand \BibitemShut  [1]{\csname bibitem#1\endcsname}%
\let\auto@bib@innerbib\@empty
\bibitem [{\citenamefont {Rubbia}(1977)}]{rubbia77}%
  \BibitemOpen
  \bibfield  {author} {\bibinfo {author} {\bibfnamefont {C.}~\bibnamefont
  {Rubbia}},\ }\bibfield  {title} {\enquote {\bibinfo {title} {{The Liquid
  Argon Time Projection Chamber: A New Concept for Neutrino Detectors}},}\
  }\href@noop {} {\bibfield  {journal} {\bibinfo  {journal} {CERN-EP-INT-77-08,
  CERN-EP-77-08}\ } (\bibinfo {year} {1977})}\BibitemShut {NoStop}%
\bibitem [{\citenamefont {Chen}\ \emph {et~al.}(1976)\citenamefont {Chen},
  \citenamefont {Condon}, \citenamefont {Barish},\ and\ \citenamefont
  {Sciulli}}]{Chen:1976pp}%
  \BibitemOpen
  \bibfield  {author} {\bibinfo {author} {\bibfnamefont {H.~H.}\ \bibnamefont
  {Chen}}, \bibinfo {author} {\bibfnamefont {P.~E.}\ \bibnamefont {Condon}},
  \bibinfo {author} {\bibfnamefont {B.~C.}\ \bibnamefont {Barish}}, \ and\
  \bibinfo {author} {\bibfnamefont {F.~J.}\ \bibnamefont {Sciulli}},\
  }\bibfield  {title} {\enquote {\bibinfo {title} {{A Neutrino detector
  sensitive to rare processes. I. A Study of neutrino electron reactions}},}\
  }\href@noop {} {\bibfield  {journal} {\bibinfo  {journal}
  {FERMILAB-PROPOSAL-0496}\ } (\bibinfo {year} {1976})}\BibitemShut {NoStop}%
\bibitem [{\citenamefont {Willis}\ and\ \citenamefont
  {Radeka}(1974)}]{willis74}%
  \BibitemOpen
  \bibfield  {author} {\bibinfo {author} {\bibfnamefont {W.J.}\ \bibnamefont
  {Willis}}\ and\ \bibinfo {author} {\bibfnamefont {V.}~\bibnamefont
  {Radeka}},\ }\bibfield  {title} {\enquote {\bibinfo {title} {{Liquid Argon
  Ionization Chambers as Total Absorption Detectors}},}\ }\href {\doibase
  10.1016/0029-554X(74)90039-1} {\bibfield  {journal} {\bibinfo  {journal}
  {Nucl. Instrum. Meth.}\ }\textbf {\bibinfo {volume} {120}},\ \bibinfo {pages}
  {221--236} (\bibinfo {year} {1974})}\BibitemShut {NoStop}%
\bibitem [{\citenamefont {Nygren}(1974)}]{Nygren:1976fe}%
  \BibitemOpen
  \bibfield  {author} {\bibinfo {author} {\bibfnamefont {D.R.}\ \bibnamefont
  {Nygren}},\ }\bibfield  {title} {\enquote {\bibinfo {title} {{The Time
  Projection Chamber: A New 4 pi Detector for Charged Particles}},}\
  }\href@noop {} {\bibfield  {journal} {\bibinfo  {journal} {eConf}\ }\textbf
  {\bibinfo {volume} {C740805}},\ \bibinfo {pages} {58} (\bibinfo {year}
  {1974})}\BibitemShut {NoStop}%
\bibitem [{\citenamefont {Acero}\ \emph {et~al.}(2019)\citenamefont {Acero}
  \emph {et~al.}}]{nova}%
  \BibitemOpen
  \bibfield  {author} {\bibinfo {author} {\bibfnamefont {M.A.}\ \bibnamefont
  {Acero}} \emph {et~al.} (\bibinfo {collaboration} {NOvA}),\ }\bibfield
  {title} {\enquote {\bibinfo {title} {{First Measurement of Neutrino
  Oscillation Parameters using Neutrinos and Antineutrinos by NOvA}},}\ }\href
  {\doibase 10.1103/PhysRevLett.123.151803} {\bibfield  {journal} {\bibinfo
  {journal} {Phys. Rev. Lett.}\ }\textbf {\bibinfo {volume} {123}},\ \bibinfo
  {pages} {151803} (\bibinfo {year} {2019})},\ \Eprint
  {http://arxiv.org/abs/1906.04907} {arXiv:1906.04907 [hep-ex]} \BibitemShut
  {NoStop}%
\bibitem [{\citenamefont {Abe}\ \emph {et~al.}(2020)\citenamefont {Abe} \emph
  {et~al.}}]{t2k}%
  \BibitemOpen
  \bibfield  {author} {\bibinfo {author} {\bibfnamefont {K.}~\bibnamefont
  {Abe}} \emph {et~al.} (\bibinfo {collaboration} {T2K}),\ }\bibfield  {title}
  {\enquote {\bibinfo {title} {{Constraint on the matter\textendash{}antimatter
  symmetry-violating phase in neutrino oscillations}},}\ }\href {\doibase
  10.1038/s41586-020-2177-0} {\bibfield  {journal} {\bibinfo  {journal}
  {Nature}\ }\textbf {\bibinfo {volume} {580}},\ \bibinfo {pages} {339--344}
  (\bibinfo {year} {2020})},\ \bibinfo {note} {[Erratum: Nature 583, E16
  (2020)]},\ \Eprint {http://arxiv.org/abs/1910.03887} {arXiv:1910.03887
  [hep-ex]} \BibitemShut {NoStop}%
\bibitem [{\citenamefont {Qian}\ and\ \citenamefont
  {Vogel}(2015)}]{Qian:2015waa}%
  \BibitemOpen
  \bibfield  {author} {\bibinfo {author} {\bibfnamefont {X.}~\bibnamefont
  {Qian}}\ and\ \bibinfo {author} {\bibfnamefont {P.}~\bibnamefont {Vogel}},\
  }\bibfield  {title} {\enquote {\bibinfo {title} {{Neutrino Mass
  Hierarchy}},}\ }\href {\doibase 10.1016/j.ppnp.2015.05.002} {\bibfield
  {journal} {\bibinfo  {journal} {Prog. Part. Nucl. Phys.}\ }\textbf {\bibinfo
  {volume} {83}},\ \bibinfo {pages} {1--30} (\bibinfo {year}
  {2015})}\BibitemShut {NoStop}%
\bibitem [{\citenamefont {Machado}\ \emph {et~al.}(2019)\citenamefont
  {Machado}, \citenamefont {Palamara},\ and\ \citenamefont
  {Schmitz}}]{Machado:2019oxb}%
  \BibitemOpen
  \bibfield  {author} {\bibinfo {author} {\bibfnamefont {Pedro~AN}\
  \bibnamefont {Machado}}, \bibinfo {author} {\bibfnamefont {Ornella}\
  \bibnamefont {Palamara}}, \ and\ \bibinfo {author} {\bibfnamefont {David~W}\
  \bibnamefont {Schmitz}},\ }\bibfield  {title} {\enquote {\bibinfo {title}
  {{The Short-Baseline Neutrino Program at Fermilab}},}\ }\href@noop {}
  {\bibfield  {journal} {\bibinfo  {journal} {Ann. Rev. Nucl. Part. Sci.}\
  }\textbf {\bibinfo {volume} {69}} (\bibinfo {year} {2019})}\BibitemShut
  {NoStop}%
\bibitem [{\citenamefont {{Particle Data Group}}(2020)}]{pdg-nn}%
  \BibitemOpen
  \bibfield  {author} {\bibinfo {author} {\bibnamefont {{Particle Data
  Group}}},\ }\bibfield  {title} {\enquote {\bibinfo {title} {{Review of
  Particle Physics: Chapter 51. Neutrino Cross Section Measurements}},}\ }\href
  {\doibase 10.1093/ptep/ptaa104} {\bibfield  {journal} {\bibinfo  {journal}
  {Progress of Theoretical and Experimental Physics}\ }\textbf {\bibinfo
  {volume} {2020}} (\bibinfo {year} {2020}),\ 10.1093/ptep/ptaa104}\BibitemShut
  {NoStop}%
\bibitem [{\citenamefont {Cavanna}\ \emph {et~al.}(2018)\citenamefont
  {Cavanna}, \citenamefont {Ereditato},\ and\ \citenamefont
  {Fleming}}]{Cavanna:2018yfk}%
  \BibitemOpen
  \bibfield  {author} {\bibinfo {author} {\bibfnamefont {F.}~\bibnamefont
  {Cavanna}}, \bibinfo {author} {\bibfnamefont {A.}~\bibnamefont {Ereditato}},
  \ and\ \bibinfo {author} {\bibfnamefont {B.~T.}\ \bibnamefont {Fleming}},\
  }\bibfield  {title} {\enquote {\bibinfo {title} {{Advances in liquid argon
  detectors}},}\ }\href {\doibase 10.1016/j.nima.2018.07.010} {\bibfield
  {journal} {\bibinfo  {journal} {Nucl. Instrum. Meth.}\ }\textbf {\bibinfo
  {volume} {A907}},\ \bibinfo {pages} {1--8} (\bibinfo {year}
  {2018})}\BibitemShut {NoStop}%
\bibitem [{\citenamefont {Amerio}\ \emph {et~al.}(2004)\citenamefont {Amerio}
  \emph {et~al.}}]{Amerio:2004ze}%
  \BibitemOpen
  \bibfield  {author} {\bibinfo {author} {\bibfnamefont {S.}~\bibnamefont
  {Amerio}} \emph {et~al.} (\bibinfo {collaboration} {ICARUS Collaboration}),\
  }\bibfield  {title} {\enquote {\bibinfo {title} {{Design, construction and
  tests of the ICARUS T600 detector}},}\ }\href {\doibase
  10.1016/j.nima.2004.02.044} {\bibfield  {journal} {\bibinfo  {journal} {Nucl.
  Instrum. Meth. A}\ }\textbf {\bibinfo {volume} {527}},\ \bibinfo {pages}
  {329--410} (\bibinfo {year} {2004})}\BibitemShut {NoStop}%
\bibitem [{\citenamefont {Anderson}\ \emph {et~al.}(2012)\citenamefont
  {Anderson} \emph {et~al.}}]{ArgoNeuT2012}%
  \BibitemOpen
  \bibfield  {author} {\bibinfo {author} {\bibfnamefont {C.}~\bibnamefont
  {Anderson}} \emph {et~al.} (\bibinfo {collaboration} {ArgoNeuT
  Collaboration}),\ }\bibfield  {title} {\enquote {\bibinfo {title} {{The
  ArgoNeuT Detector in the NuMI Low-Energy beam line at Fermilab}},}\ }\href
  {\doibase 10.1088/1748-0221/7/10/P10019} {\bibfield  {journal} {\bibinfo
  {journal} {JINST}\ }\textbf {\bibinfo {volume} {7}},\ \bibinfo {pages}
  {P10019} (\bibinfo {year} {2012})}\BibitemShut {NoStop}%
\bibitem [{\citenamefont {Acciarri}\ \emph
  {et~al.}(2017{\natexlab{a}})\citenamefont {Acciarri} \emph
  {et~al.}}]{Acciarri:2016smi}%
  \BibitemOpen
  \bibfield  {author} {\bibinfo {author} {\bibfnamefont {R.}~\bibnamefont
  {Acciarri}} \emph {et~al.} (\bibinfo {collaboration} {MicroBooNE
  Collaboration}),\ }\bibfield  {title} {\enquote {\bibinfo {title} {{Design
  and Construction of the MicroBooNE Detector}},}\ }\href {\doibase
  10.1088/1748-0221/12/02/P02017} {\bibfield  {journal} {\bibinfo  {journal}
  {JINST}\ }\textbf {\bibinfo {volume} {12}},\ \bibinfo {pages} {P02017}
  (\bibinfo {year} {2017}{\natexlab{a}})}\BibitemShut {NoStop}%
\bibitem [{\citenamefont {Badhrees}\ \emph {et~al.}(2012)\citenamefont
  {Badhrees} \emph {et~al.}}]{Badhrees:2012zz}%
  \BibitemOpen
  \bibfield  {author} {\bibinfo {author} {\bibfnamefont {I.}~\bibnamefont
  {Badhrees}} \emph {et~al.},\ }\bibfield  {title} {\enquote {\bibinfo {title}
  {{Argontube: An R\&D liquid Argon Time Projection Chamber}},}\ }\href
  {\doibase 10.1088/1748-0221/7/02/C02011} {\bibfield  {journal} {\bibinfo
  {journal} {JINST}\ }\textbf {\bibinfo {volume} {7}},\ \bibinfo {pages}
  {C02011} (\bibinfo {year} {2012})}\BibitemShut {NoStop}%
\bibitem [{\citenamefont {Bhandari}\ \emph {et~al.}(2019)\citenamefont
  {Bhandari} \emph {et~al.}}]{Bhandari:2019rat}%
  \BibitemOpen
  \bibfield  {author} {\bibinfo {author} {\bibfnamefont {B.}~\bibnamefont
  {Bhandari}} \emph {et~al.} (\bibinfo {collaboration} {CAPTAIN
  Collaboration}),\ }\bibfield  {title} {\enquote {\bibinfo {title} {{First
  Measurement of the Total Neutron Cross Section on Argon Between 100 and 800
  MeV}},}\ }\href {\doibase 10.1103/PhysRevLett.123.042502} {\bibfield
  {journal} {\bibinfo  {journal} {Phys. Rev. Lett.}\ }\textbf {\bibinfo
  {volume} {123}},\ \bibinfo {pages} {042502} (\bibinfo {year}
  {2019})}\BibitemShut {NoStop}%
\bibitem [{\citenamefont {Hahn}\ \emph {et~al.}(2016)\citenamefont {Hahn} \emph
  {et~al.}}]{Hahn:2016tia}%
  \BibitemOpen
  \bibfield  {author} {\bibinfo {author} {\bibfnamefont {Alan}\ \bibnamefont
  {Hahn}} \emph {et~al.} (\bibinfo {collaboration} {LBNE Collaboration}),\
  }\bibfield  {title} {\enquote {\bibinfo {title} {{The LBNE 35 Ton Prototype
  Cryostat}},}\ }in\ \href {\doibase 10.1109/NSSMIC.2014.7431158} {\emph
  {\bibinfo {booktitle} {{2014 IEEE Nuclear Science Symposium and Medical
  Imaging Conference and 21st Symposium on Room-Temperature Semiconductor X-ray
  and Gamma-ray Detectors}}}}\ (\bibinfo {year} {2016})\ p.\ \bibinfo {pages}
  {7431158}\BibitemShut {NoStop}%
\bibitem [{\citenamefont {Paley}\ \emph {et~al.}(2014)\citenamefont {Paley}
  \emph {et~al.}}]{Cavanna:2014iqa}%
  \BibitemOpen
  \bibfield  {author} {\bibinfo {author} {\bibfnamefont {J.}~\bibnamefont
  {Paley}} \emph {et~al.} (\bibinfo {collaboration} {LArIAT Collaboration}),\
  }\href@noop {} {\enquote {\bibinfo {title} {{LArIAT: Liquid Argon In A
  Testbeam}},}\ } (\bibinfo {year} {2014}),\ \Eprint
  {http://arxiv.org/abs/1406.5560} {arXiv:1406.5560 [physics.ins-det]}
  \BibitemShut {NoStop}%
\bibitem [{\citenamefont {Abi}\ \emph {et~al.}(2020{\natexlab{a}})\citenamefont
  {Abi} \emph {et~al.}}]{Abi:2020mwi}%
  \BibitemOpen
  \bibfield  {author} {\bibinfo {author} {\bibfnamefont {B.}~\bibnamefont
  {Abi}} \emph {et~al.} (\bibinfo {collaboration} {DUNE Collaboration}),\
  }\bibfield  {title} {\enquote {\bibinfo {title} {{First results on
  ProtoDUNE-SP liquid argon time projection chamber performance from a beam
  test at the CERN Neutrino Platform}},}\ }\href {\doibase
  10.1088/1748-0221/15/12/P12004} {\bibfield  {journal} {\bibinfo  {journal}
  {JINST}\ }\textbf {\bibinfo {volume} {15}},\ \bibinfo {pages} {P12004}
  (\bibinfo {year} {2020}{\natexlab{a}})},\ \Eprint
  {http://arxiv.org/abs/2007.06722} {arXiv:2007.06722 [physics.ins-det]}
  \BibitemShut {NoStop}%
\bibitem [{\citenamefont {Antonello}\ \emph {et~al.}(2015)\citenamefont
  {Antonello} \emph {et~al.}}]{Antonello:2015lea}%
  \BibitemOpen
  \bibfield  {author} {\bibinfo {author} {\bibfnamefont {M.}~\bibnamefont
  {Antonello}} \emph {et~al.} (\bibinfo {collaboration} {MicroBooNE, LAr1-ND,
  and ICARUS-WA104 Collaboration}),\ }\href@noop {} {\enquote {\bibinfo {title}
  {{A Proposal for a Three Detector Short-Baseline Neutrino Oscillation Program
  in the Fermilab Booster Neutrino Beam}},}\ } (\bibinfo {year} {2015}),\
  \Eprint {http://arxiv.org/abs/1503.01520} {arXiv:1503.01520
  [physics.ins-det]} \BibitemShut {NoStop}%
\bibitem [{\citenamefont {Abi}\ \emph {et~al.}(2020{\natexlab{b}})\citenamefont
  {Abi} \emph {et~al.}}]{dune-tdr-1}%
  \BibitemOpen
  \bibfield  {author} {\bibinfo {author} {\bibfnamefont {Babak}\ \bibnamefont
  {Abi}} \emph {et~al.} (\bibinfo {collaboration} {DUNE Collaboration}),\
  }\bibfield  {title} {\enquote {\bibinfo {title} {{Deep Underground Neutrino
  Experiment (DUNE), Far Detector Technical Design Report, Volume I
  Introduction to DUNE}},}\ }\href {\doibase 10.1088/1748-0221/15/08/T08008}
  {\bibfield  {journal} {\bibinfo  {journal} {JINST}\ }\textbf {\bibinfo
  {volume} {15}},\ \bibinfo {pages} {T08008} (\bibinfo {year}
  {2020}{\natexlab{b}})}\BibitemShut {NoStop}%
\bibitem [{\citenamefont {Aguilar-Arevalo}\ \emph {et~al.}(2009)\citenamefont
  {Aguilar-Arevalo} \emph {et~al.}}]{AguilarArevalo:2008yp}%
  \BibitemOpen
  \bibfield  {author} {\bibinfo {author} {\bibfnamefont {A.~A.}\ \bibnamefont
  {Aguilar-Arevalo}} \emph {et~al.} (\bibinfo {collaboration} {MiniBooNE
  Collaboration}),\ }\bibfield  {title} {\enquote {\bibinfo {title} {{The
  Neutrino Flux prediction at MiniBooNE}},}\ }\href {\doibase
  10.1103/PhysRevD.79.072002} {\bibfield  {journal} {\bibinfo  {journal} {Phys.
  Rev. D.}\ }\textbf {\bibinfo {volume} {79}},\ \bibinfo {pages} {072002}
  (\bibinfo {year} {2009})}\BibitemShut {NoStop}%
\bibitem [{\citenamefont {Briese}\ \emph {et~al.}(2013)\citenamefont {Briese}
  \emph {et~al.}}]{Briese:2013wua}%
  \BibitemOpen
  \bibfield  {author} {\bibinfo {author} {\bibfnamefont {T.}~\bibnamefont
  {Briese}} \emph {et~al.},\ }\bibfield  {title} {\enquote {\bibinfo {title}
  {{Testing of Cryogenic Photomultiplier Tubes for the MicroBooNE
  Experiment}},}\ }\href {\doibase 10.1088/1748-0221/8/07/T07005} {\bibfield
  {journal} {\bibinfo  {journal} {JINST}\ }\textbf {\bibinfo {volume} {8}},\
  \bibinfo {pages} {T07005} (\bibinfo {year} {2013})},\ \Eprint
  {http://arxiv.org/abs/1304.0821} {arXiv:1304.0821 [physics.ins-det]}
  \BibitemShut {NoStop}%
\bibitem [{\citenamefont {Radeka}\ \emph {et~al.}(2011)\citenamefont {Radeka}
  \emph {et~al.}}]{Radeka:2011zz}%
  \BibitemOpen
  \bibfield  {author} {\bibinfo {author} {\bibfnamefont {Veljko}\ \bibnamefont
  {Radeka}} \emph {et~al.},\ }\bibfield  {title} {\enquote {\bibinfo {title}
  {{Cold electronics for 'Giant' Liquid Argon Time Projection Chambers}},}\
  }\bibfield  {booktitle} {\emph {\bibinfo {booktitle} {{Giant liquid argon
  charge imaging experiment. Proceedings, 1st International Workshop, GLA2010,
  Tsukuba, Japan, March 29-31, 2010}}},\ }\href {\doibase
  10.1088/1742-6596/308/1/012021} {\bibfield  {journal} {\bibinfo  {journal}
  {J. Phys. Conf. Ser.}\ }\textbf {\bibinfo {volume} {308}},\ \bibinfo {pages}
  {012021} (\bibinfo {year} {2011})}\BibitemShut {NoStop}%
\bibitem [{\citenamefont {Acciarri}\ \emph
  {et~al.}(2017{\natexlab{b}})\citenamefont {Acciarri} \emph
  {et~al.}}]{Acciarri:2017rnj}%
  \BibitemOpen
  \bibfield  {author} {\bibinfo {author} {\bibfnamefont {R.}~\bibnamefont
  {Acciarri}} \emph {et~al.} (\bibinfo {collaboration} {MicroBooNE
  Collaboration}),\ }\bibfield  {title} {\enquote {\bibinfo {title}
  {{Measurement of cosmic-ray reconstruction efficiencies in the MicroBooNE
  LArTPC using a small external cosmic-ray counter}},}\ }\href {\doibase
  10.1088/1748-0221/12/12/P12030} {\bibfield  {journal} {\bibinfo  {journal}
  {JINST}\ }\textbf {\bibinfo {volume} {12}},\ \bibinfo {pages} {P12030}
  (\bibinfo {year} {2017}{\natexlab{b}})}\BibitemShut {NoStop}%
\bibitem [{\citenamefont {Adams}\ \emph
  {et~al.}(2019{\natexlab{a}})\citenamefont {Adams} \emph
  {et~al.}}]{Adams:2018fud}%
  \BibitemOpen
  \bibfield  {author} {\bibinfo {author} {\bibfnamefont {C.}~\bibnamefont
  {Adams}} \emph {et~al.} (\bibinfo {collaboration} {MicroBooNE
  Collaboration}),\ }\bibfield  {title} {\enquote {\bibinfo {title}
  {{Comparison of $\nu_\mu$-Ar multiplicity distributions observed by
  MicroBooNE to GENIE model predictions}},}\ }\href {\doibase
  10.1140/epjc/s10052-019-6742-3} {\bibfield  {journal} {\bibinfo  {journal}
  {Eur. Phys. J.}\ }\textbf {\bibinfo {volume} {C79}},\ \bibinfo {pages} {248}
  (\bibinfo {year} {2019}{\natexlab{a}})}\BibitemShut {NoStop}%
\bibitem [{\citenamefont {Adams}\ \emph
  {et~al.}(2019{\natexlab{b}})\citenamefont {Adams} \emph
  {et~al.}}]{Adams:2018sgn}%
  \BibitemOpen
  \bibfield  {author} {\bibinfo {author} {\bibfnamefont {C.}~\bibnamefont
  {Adams}} \emph {et~al.} (\bibinfo {collaboration} {MicroBooNE
  Collaboration}),\ }\bibfield  {title} {\enquote {\bibinfo {title} {{First
  measurement of $\nu_{\mu}$ charged-current $\pi^{0}$ production on argon with
  the MicroBooNE detector}},}\ }\href {\doibase 10.1103/PhysRevD.99.091102}
  {\bibfield  {journal} {\bibinfo  {journal} {Phys. Rev.}\ }\textbf {\bibinfo
  {volume} {D99}},\ \bibinfo {pages} {091102} (\bibinfo {year}
  {2019}{\natexlab{b}})}\BibitemShut {NoStop}%
\bibitem [{\citenamefont {Adams}\ \emph
  {et~al.}(2019{\natexlab{c}})\citenamefont {Adams} \emph
  {et~al.}}]{Adams:2018lzd}%
  \BibitemOpen
  \bibfield  {author} {\bibinfo {author} {\bibfnamefont {C.}~\bibnamefont
  {Adams}} \emph {et~al.} (\bibinfo {collaboration} {MicroBooNE
  Collaboration}),\ }\bibfield  {title} {\enquote {\bibinfo {title} {{Rejecting
  cosmic background for exclusive charged current quasi elastic neutrino
  interaction studies with Liquid Argon TPCs; a case study with the MicroBooNE
  detector}},}\ }\href {\doibase 10.1140/epjc/s10052-019-7184-7} {\bibfield
  {journal} {\bibinfo  {journal} {Eur. Phys. J.}\ }\textbf {\bibinfo {volume}
  {C79}},\ \bibinfo {pages} {673} (\bibinfo {year}
  {2019}{\natexlab{c}})}\BibitemShut {NoStop}%
\bibitem [{\citenamefont {Abratenko}\ \emph {et~al.}(2019)\citenamefont
  {Abratenko} \emph {et~al.}}]{Adams:2019iqc}%
  \BibitemOpen
  \bibfield  {author} {\bibinfo {author} {\bibfnamefont {P.}~\bibnamefont
  {Abratenko}} \emph {et~al.} (\bibinfo {collaboration} {MicroBooNE
  Collaboration}),\ }\bibfield  {title} {\enquote {\bibinfo {title} {{First
  Measurement of Inclusive Muon Neutrino Charged Current Differential Cross
  Sections on Argon at $E_\nu\sim$0.8 GeV with the MicroBooNE Detector}},}\
  }\href {\doibase 10.1103/PhysRevLett.123.131801} {\bibfield  {journal}
  {\bibinfo  {journal} {Phys. Rev. Lett.}\ }\textbf {\bibinfo {volume} {123}},\
  \bibinfo {pages} {131801} (\bibinfo {year} {2019})}\BibitemShut {NoStop}%
\bibitem [{\citenamefont {Acciarri}\ \emph
  {et~al.}(2017{\natexlab{c}})\citenamefont {Acciarri} \emph
  {et~al.}}]{Acciarri:2017sde}%
  \BibitemOpen
  \bibfield  {author} {\bibinfo {author} {\bibfnamefont {R.}~\bibnamefont
  {Acciarri}} \emph {et~al.} (\bibinfo {collaboration} {MicroBooNE
  Collaboration}),\ }\bibfield  {title} {\enquote {\bibinfo {title} {{Noise
  Characterization and Filtering in the MicroBooNE Liquid Argon TPC}},}\ }\href
  {\doibase 10.1088/1748-0221/12/08/P08003} {\bibfield  {journal} {\bibinfo
  {journal} {JINST}\ }\textbf {\bibinfo {volume} {12}},\ \bibinfo {pages}
  {P08003} (\bibinfo {year} {2017}{\natexlab{c}})}\BibitemShut {NoStop}%
\bibitem [{\citenamefont {Adams}\ \emph
  {et~al.}(2018{\natexlab{a}})\citenamefont {Adams} \emph
  {et~al.}}]{Adams:2018dra}%
  \BibitemOpen
  \bibfield  {author} {\bibinfo {author} {\bibfnamefont {C.}~\bibnamefont
  {Adams}} \emph {et~al.} (\bibinfo {collaboration} {MicroBooNE
  Collaboration}),\ }\bibfield  {title} {\enquote {\bibinfo {title}
  {{Ionization electron signal processing in single phase LArTPCs. Part I.
  Algorithm Description and quantitative evaluation with MicroBooNE
  simulation}},}\ }\href {\doibase 10.1088/1748-0221/13/07/P07006} {\bibfield
  {journal} {\bibinfo  {journal} {JINST}\ }\textbf {\bibinfo {volume} {13}},\
  \bibinfo {pages} {P07006--P07006} (\bibinfo {year}
  {2018}{\natexlab{a}})}\BibitemShut {NoStop}%
\bibitem [{\citenamefont {Baller}(2017)}]{Baller:2017ugz}%
  \BibitemOpen
  \bibfield  {author} {\bibinfo {author} {\bibfnamefont {Bruce}\ \bibnamefont
  {Baller}},\ }\bibfield  {title} {\enquote {\bibinfo {title} {{Liquid argon
  TPC signal formation, signal processing and reconstruction techniques}},}\
  }\href {\doibase 10.1088/1748-0221/12/07/P07010} {\bibfield  {journal}
  {\bibinfo  {journal} {JINST}\ }\textbf {\bibinfo {volume} {12}},\ \bibinfo
  {pages} {P07010} (\bibinfo {year} {2017})}\BibitemShut {NoStop}%
\bibitem [{\citenamefont {Adams}\ \emph
  {et~al.}(2018{\natexlab{b}})\citenamefont {Adams} \emph
  {et~al.}}]{Adams:2018gbi}%
  \BibitemOpen
  \bibfield  {author} {\bibinfo {author} {\bibfnamefont {C.}~\bibnamefont
  {Adams}} \emph {et~al.} (\bibinfo {collaboration} {MicroBooNE
  Collaboration}),\ }\bibfield  {title} {\enquote {\bibinfo {title}
  {{Ionization electron signal processing in single phase LArTPCs. Part II.
  Data/simulation comparison and performance in MicroBooNE}},}\ }\href
  {\doibase 10.1088/1748-0221/13/07/P07007} {\bibfield  {journal} {\bibinfo
  {journal} {JINST}\ }\textbf {\bibinfo {volume} {13}},\ \bibinfo {pages}
  {P07007} (\bibinfo {year} {2018}{\natexlab{b}})}\BibitemShut {NoStop}%
\bibitem [{\citenamefont {Qian}\ \emph {et~al.}(2018)\citenamefont {Qian},
  \citenamefont {Zhang}, \citenamefont {Viren},\ and\ \citenamefont
  {Diwan}}]{Qian:2018qbv}%
  \BibitemOpen
  \bibfield  {author} {\bibinfo {author} {\bibfnamefont {Xin}\ \bibnamefont
  {Qian}}, \bibinfo {author} {\bibfnamefont {Chao}\ \bibnamefont {Zhang}},
  \bibinfo {author} {\bibfnamefont {Brett}\ \bibnamefont {Viren}}, \ and\
  \bibinfo {author} {\bibfnamefont {Milind}\ \bibnamefont {Diwan}},\ }\bibfield
   {title} {\enquote {\bibinfo {title} {{Three-dimensional Imaging for Large
  LArTPCs}},}\ }\href {\doibase 10.1088/1748-0221/13/05/P05032} {\bibfield
  {journal} {\bibinfo  {journal} {JINST}\ }\textbf {\bibinfo {volume} {13}},\
  \bibinfo {pages} {P05032} (\bibinfo {year} {2018})}\BibitemShut {NoStop}%
\bibitem [{\citenamefont {Abratenko}\ \emph
  {et~al.}(2020{\natexlab{a}})\citenamefont {Abratenko} \emph
  {et~al.}}]{wire-cell-uboone}%
  \BibitemOpen
  \bibfield  {author} {\bibinfo {author} {\bibfnamefont {P.}~\bibnamefont
  {Abratenko}} \emph {et~al.} (\bibinfo {collaboration} {MicroBooNE}),\
  }\bibfield  {title} {\enquote {\bibinfo {title} {{Neutrino Event Selection in
  the MicroBooNE Liquid Argon Time Projection Chamber using Wire-Cell 3-D
  Imaging, Clustering, and Charge-Light Matching}},}\ }\href@noop {} {\
  (\bibinfo {year} {2020}{\natexlab{a}})},\ \bibinfo {note} {accepted by
  JINST},\ \Eprint {http://arxiv.org/abs/2011.01375} {arXiv:2011.01375
  [physics.ins-det]} \BibitemShut {NoStop}%
\bibitem [{\citenamefont {Adamson}\ \emph {et~al.}(2016)\citenamefont {Adamson}
  \emph {et~al.}}]{Adamson:2016xxw}%
  \BibitemOpen
  \bibfield  {author} {\bibinfo {author} {\bibfnamefont {P.}~\bibnamefont
  {Adamson}} \emph {et~al.} (\bibinfo {collaboration} {NOvA Collaboration}),\
  }\bibfield  {title} {\enquote {\bibinfo {title} {{First measurement of
  muon-neutrino disappearance in NOvA}},}\ }\href {\doibase
  10.1103/PhysRevD.93.051104} {\bibfield  {journal} {\bibinfo  {journal} {Phys.
  Rev.}\ }\textbf {\bibinfo {volume} {D93}},\ \bibinfo {pages} {051104}
  (\bibinfo {year} {2016})}\BibitemShut {NoStop}%
\bibitem [{\citenamefont {Aliaga}\ \emph {et~al.}(2014)\citenamefont {Aliaga}
  \emph {et~al.}}]{Aliaga:2013uqz}%
  \BibitemOpen
  \bibfield  {author} {\bibinfo {author} {\bibfnamefont {L.}~\bibnamefont
  {Aliaga}} \emph {et~al.} (\bibinfo {collaboration} {MINERvA Collaboration}),\
  }\bibfield  {title} {\enquote {\bibinfo {title} {{Design, Calibration, and
  Performance of the MINERvA Detector}},}\ }\href {\doibase
  10.1016/j.nima.2013.12.053} {\bibfield  {journal} {\bibinfo  {journal} {Nucl.
  Instrum. Meth.}\ }\textbf {\bibinfo {volume} {A743}},\ \bibinfo {pages}
  {130--159} (\bibinfo {year} {2014})}\BibitemShut {NoStop}%
\bibitem [{\citenamefont {Michael}\ \emph {et~al.}(2008)\citenamefont {Michael}
  \emph {et~al.}}]{Michael:2008bc}%
  \BibitemOpen
  \bibfield  {author} {\bibinfo {author} {\bibfnamefont {D.~G.}\ \bibnamefont
  {Michael}} \emph {et~al.} (\bibinfo {collaboration} {MINOS Collaboration}),\
  }\bibfield  {title} {\enquote {\bibinfo {title} {{The Magnetized steel and
  scintillator calorimeters of the MINOS experiment}},}\ }\href {\doibase
  10.1016/j.nima.2008.08.003} {\bibfield  {journal} {\bibinfo  {journal} {Nucl.
  Instrum. Meth.}\ }\textbf {\bibinfo {volume} {A596}},\ \bibinfo {pages}
  {190--228} (\bibinfo {year} {2008})}\BibitemShut {NoStop}%
\bibitem [{\citenamefont {Agostinelli}\ \emph {et~al.}(2003)\citenamefont
  {Agostinelli} \emph {et~al.}}]{geant}%
  \BibitemOpen
  \bibfield  {author} {\bibinfo {author} {\bibfnamefont {S.}~\bibnamefont
  {Agostinelli}} \emph {et~al.},\ }\bibfield  {title} {\enquote {\bibinfo
  {title} {{GEAnt4--a simulation toolkit}},}\ }\href@noop {} {\bibfield
  {journal} {\bibinfo  {journal} {Nucl. Instrum. Methods Phys. Res.}\ }\textbf
  {\bibinfo {volume} {Sect. A 506}},\ \bibinfo {pages} {250} (\bibinfo {year}
  {2003})}\BibitemShut {NoStop}%
\bibitem [{\citenamefont {Allison}\ \emph {et~al.}(2006)\citenamefont {Allison}
  \emph {et~al.}}]{geant_2}%
  \BibitemOpen
  \bibfield  {author} {\bibinfo {author} {\bibfnamefont {J.}~\bibnamefont
  {Allison}} \emph {et~al.},\ }\bibfield  {title} {\enquote {\bibinfo {title}
  {{Geant4 developments and applications}},}\ }\href@noop {} {\bibfield
  {journal} {\bibinfo  {journal} {IEEE Trans. Nucl. Sci.}\ }\textbf {\bibinfo
  {volume} {53}},\ \bibinfo {pages} {270} (\bibinfo {year} {2006})}\BibitemShut
  {NoStop}%
\bibitem [{\citenamefont {Cand\`{e}s}\ \emph {et~al.}(2006)\citenamefont
  {Cand\`{e}s}, \citenamefont {Romberg},\ and\ \citenamefont {Tao}}]{cs}%
  \BibitemOpen
  \bibfield  {author} {\bibinfo {author} {\bibfnamefont {E.~J.}\ \bibnamefont
  {Cand\`{e}s}}, \bibinfo {author} {\bibfnamefont {J.~K.}\ \bibnamefont
  {Romberg}}, \ and\ \bibinfo {author} {\bibfnamefont {T.}~\bibnamefont
  {Tao}},\ }\bibfield  {title} {\enquote {\bibinfo {title} {Stable signal
  recovery from incomplete and inaccurate measurements},}\ }\href {\doibase
  10.1002/cpa.20124} {\bibfield  {journal} {\bibinfo  {journal} {Communications
  on Pure and Applied Mathematics}\ }\textbf {\bibinfo {volume} {59}},\
  \bibinfo {pages} {1207--1223} (\bibinfo {year} {2006})}\BibitemShut {NoStop}%
\bibitem [{\citenamefont {Abratenko}\ \emph {et~al.}(2021)\citenamefont
  {Abratenko} \emph {et~al.}}]{wire-cell-generic-neutrino}%
  \BibitemOpen
  \bibfield  {author} {\bibinfo {author} {\bibfnamefont {P.}~\bibnamefont
  {Abratenko}} \emph {et~al.} (\bibinfo {collaboration} {MicroBooNE}),\
  }\bibfield  {title} {\enquote {\bibinfo {title} {{Cosmic Ray Background
  Rejection with Wire-Cell LArTPC Event Reconstruction in the MicroBooNE
  Detector}},}\ }\href {\doibase 10.1103/PhysRevApplied.15.064071} {\bibfield
  {journal} {\bibinfo  {journal} {Phys. Rev. Applied}\ }\textbf {\bibinfo
  {volume} {15}},\ \bibinfo {pages} {064071} (\bibinfo {year} {2021})},\
  \Eprint {http://arxiv.org/abs/2101.05076} {arXiv:2101.05076
  [physics.ins-det]} \BibitemShut {NoStop}%
\bibitem [{\citenamefont {Adams}\ \emph {et~al.}(2020)\citenamefont {Adams}
  \emph {et~al.}}]{Adams:2019qrr}%
  \BibitemOpen
  \bibfield  {author} {\bibinfo {author} {\bibfnamefont {C.}~\bibnamefont
  {Adams}} \emph {et~al.} (\bibinfo {collaboration} {MicroBooNE
  Collaboration}),\ }\bibfield  {title} {\enquote {\bibinfo {title} {{A Method
  to Determine the Electric Field of Liquid Argon Time Projection Chambers
  Using a UV Laser System and its Application in MicroBooNE}},}\ }\href
  {\doibase 10.1088/1748-0221/15/07/P07010} {\bibfield  {journal} {\bibinfo
  {journal} {JINST}\ }\textbf {\bibinfo {volume} {15}},\ \bibinfo {pages}
  {P07010} (\bibinfo {year} {2020})}\BibitemShut {NoStop}%
\bibitem [{\citenamefont {Abratenko}\ \emph
  {et~al.}(2020{\natexlab{b}})\citenamefont {Abratenko} \emph
  {et~al.}}]{Abratenko:2020bbx}%
  \BibitemOpen
  \bibfield  {author} {\bibinfo {author} {\bibfnamefont {P.}~\bibnamefont
  {Abratenko}} \emph {et~al.} (\bibinfo {collaboration} {MicroBooNE}),\
  }\bibfield  {title} {\enquote {\bibinfo {title} {{Measurement of space charge
  effects in the MicroBooNE LArTPC using cosmic muons}},}\ }\href {\doibase
  10.1088/1748-0221/15/12/P12037} {\bibfield  {journal} {\bibinfo  {journal}
  {JINST}\ }\textbf {\bibinfo {volume} {15}},\ \bibinfo {pages} {P12037}
  (\bibinfo {year} {2020}{\natexlab{b}})},\ \Eprint
  {http://arxiv.org/abs/2008.09765} {arXiv:2008.09765 [physics.ins-det]}
  \BibitemShut {NoStop}%
\bibitem [{\citenamefont {Antonello}\ \emph {et~al.}(2013)\citenamefont
  {Antonello} \emph {et~al.}}]{antonello:2012hu}%
  \BibitemOpen
  \bibfield  {author} {\bibinfo {author} {\bibfnamefont {M.}~\bibnamefont
  {Antonello}} \emph {et~al.},\ }\bibfield  {title} {\enquote {\bibinfo {title}
  {{Precise 3D track reconstruction algorithm for the ICARUS T600 liquid argon
  time projection chamber detector}},}\ }\href {\doibase 10.1155/2013/260820}
  {\bibfield  {journal} {\bibinfo  {journal} {Adv. High Energy Phys.}\ }\textbf
  {\bibinfo {volume} {2013}},\ \bibinfo {pages} {260820} (\bibinfo {year}
  {2013})}\BibitemShut {NoStop}%
\bibitem [{Ste(2019)}]{Steiner}%
  \BibitemOpen
  \href@noop {} {\enquote {\bibinfo {title} {{Steiner tree greedy
  algorithm}},}\ }\bibinfo {howpublished}
  {\url{http://paal.mimuw.edu.pl/docs/index.html}} (\bibinfo {year}
  {2019})\BibitemShut {NoStop}%
\bibitem [{BiC(2019)}]{BiCGSTAB}%
  \BibitemOpen
  \href@noop {} {\enquote {\bibinfo {title} {{Biconjugate gradient stablized
  method (BiCGSTAB)}},}\ }\bibinfo {howpublished}
  {\url{https://eigen.tuxfamily.org/dox/classEigen_1_1BiCGSTAB.html}} (\bibinfo
  {year} {2019})\BibitemShut {NoStop}%
\bibitem [{\citenamefont {Kolmogorov}(1933)}]{ks_test1}%
  \BibitemOpen
  \bibfield  {author} {\bibinfo {author} {\bibfnamefont {A.}~\bibnamefont
  {Kolmogorov}},\ }\bibfield  {title} {\enquote {\bibinfo {title} {{Sulla
  determinazione empirica di una legge di distribuzione}},}\ }\href@noop {}
  {\bibfield  {journal} {\bibinfo  {journal} {G. Ist. Ital. Attuari}\ }\textbf
  {\bibinfo {volume} {{\bf 4}}},\ \bibinfo {pages} {83} (\bibinfo {year}
  {1933})}\BibitemShut {NoStop}%
\bibitem [{\citenamefont {Adams}\ \emph
  {et~al.}(2019{\natexlab{d}})\citenamefont {Adams} \emph
  {et~al.}}]{Adams:2019bzt}%
  \BibitemOpen
  \bibfield  {author} {\bibinfo {author} {\bibfnamefont {C.}~\bibnamefont
  {Adams}} \emph {et~al.} (\bibinfo {collaboration} {MicroBooNE
  Collaboration}),\ }\bibfield  {title} {\enquote {\bibinfo {title} {{Design
  and construction of the MicroBooNE Cosmic Ray Tagger system}},}\ }\href
  {\doibase 10.1088/1748-0221/14/04/P04004} {\bibfield  {journal} {\bibinfo
  {journal} {JINST}\ }\textbf {\bibinfo {volume} {14}},\ \bibinfo {pages}
  {P04004} (\bibinfo {year} {2019}{\natexlab{d}})}\BibitemShut {NoStop}%
\bibitem [{\citenamefont {Aguilar-Arevalo}\ \emph {et~al.}(2020)\citenamefont
  {Aguilar-Arevalo} \emph {et~al.}}]{Aguilar-Arevalo:2020nvw}%
  \BibitemOpen
  \bibfield  {author} {\bibinfo {author} {\bibfnamefont {A.~A.}\ \bibnamefont
  {Aguilar-Arevalo}} \emph {et~al.} (\bibinfo {collaboration} {MiniBooNE}),\
  }\bibfield  {title} {\enquote {\bibinfo {title} {{Updated MiniBooNE Neutrino
  Oscillation Results with Increased Data and New Background Studies}},}\
  }\href@noop {} {\  (\bibinfo {year} {2020})},\ \Eprint
  {http://arxiv.org/abs/2006.16883} {arXiv:2006.16883 [hep-ex]} \BibitemShut
  {NoStop}%
\end{thebibliography}%
	
\end{document}